# Investigation of Mist and Air Film Cooling in a Two-Phase Rotating Detonation Combustor with Liquid Kerosene

Yeqi Zhou[1], Songbai Yao[1,2,*], Wenwu Zhang[1,2]

[1]Zhejiang Key Laboratory of Laser Extreme Manufacturing for Difficult-to-Machine Materials, Ningbo Institute of Materials Technology and Engineering, Chinese Academy of Sciences, Ningbo 315201, China

[2]University of Chinese Academy of Sciences, Beijing 100049, China

**Abstract**

We present a numerical investigation of kerosene droplet mist film cooling for the thermal protection of the rotating detonation combustor (RDC) and compare its performance with conventional air film cooling and combined mist/air cooling scheme. In the study, the cooling behavior of kerosene droplets injected through wall film holes is numerically examined and compared with air film cooling and a combined mist/air cooling strategy, building on a benchmark validation against flat-plate experimental data. The results show that air film cooling exhibits an optimal operating range, beyond which excessive injection degrades film stability due to strong interaction with the rotating detonation wave. In contrast, kerosene-based mist cooling forms a more persistent near-wall cooling layer, providing enhanced heat removal through phase change and exhibiting improved resistance to film separation. In mist cooling, the droplet size primarily affects the immediate downstream cooling performance, with intermediate-sized droplets offering the improved balance between evaporation rate and film continuity. A combined mist/air cooling scheme can further improve cooling efficiency and accelerate wall temperature recovery after detonation wave passage while maintaining moderate impacts on the mainstream flow. Additionally, although kerosene droplets partially participate in combustion under film hole injection, the associated thermal load does not offset the overall cooling benefit. These findings demonstrate the feasibility and advantages of kerosene-based cooling schemes for RDC thermal management.

## 1. Introduction

Rotating Detonation Combustors (RDCs) have broad application prospects in aerospace propulsion and ground power fields due to their high energy release efficiency and compact structure, making them a research spotlight in the combustion field in recent years [1-9]. However, the extremely high temperature and intense heat flux impact generated by detonation waves inside RDCs impose strict requirements on the thermal protection performance of the combustor wall [10-17]. Without effective cooling measures, wall materials are prone to failure due to high temperatures, which severely restricts the long-term stable operation and engineering application of RDCs.

*Corresponding author: yaosongbai@nimte.ac.cn (S. Yao)

Consequently, effective implementation of cooling technologies for RDCs has been the subject of recent investigation [18-22].

As an efficient active thermal protection method, film cooling [23, 24] injects low-temperature coolant into the flow field through film holes on the wall, forming a low-temperature protective film on the wall, which can significantly reduce the wall heat flux density and temperature. Therefore, it is widely used for the thermal protection of hot section components of conventional combustion engines [25-31]. Recently, there has been some progress in the application of film cooling in RDCs. For example, Yu et al. [32, 33] explored the performance of combustor wall film cooling in a hydrogen-air RDC through a series of numerical simulations and experiments. In these studies, both conventional cylindrical and advanced complex-shaped film-cooling hole configurations were examined and compared. In the numerical studies, Tian et al. [34, 35] investigated a hydrogen–air RDC model featuring cylindrical film-cooling holes on the combustor wall as well as an annular slit injection structure. Using high-fidelity LES simulations, Sridhara et al. [36] observed secondary combustion when the cooling air interacted with unburned hydrogen, which increased the combustion efficiency from 90.5% to 96.8% but also intensified local heat release. Overall, the application of film cooling reduced RDC wall temperatures by approximately 500 K. Additionally, film cooling has also been investigated for the thermal protection of the nozzle [37] and turbines [38, 39] of RDCs.

However, significant research gaps and key issues regarding film cooling in RDCs remain to be addressed. A key deficiency in the existing literature is that most previous studies consider only gaseous-fuel RDCs and use gaseous air or nitrogen as the coolant. In reality, RDCs operate across diverse application scenarios, where the availability of coolant and system constraints vary considerably. For instance, in rocket-based RDCs, a dedicated cooling air supply is typically unavailable. In air-breathing configurations such as ramjets, diverting a substantial portion of inlet air for wall cooling may degrade propulsion performance. Therefore, it is important to investigate alternative coolants for RDC film cooling. Liquid fuels are viable option due to their inherent cooling potential and compatibility with combustion processes [40]. For example, Kirchberger et al. [41] compared kerosene and nitrogen as coolants in a kerosene rocket engine, and found that kerosene had higher cooling efficiency and exhibited a significant cooling effect over the entire range, while nitrogen had limited cooling efficiency and only worked effectively over a short distance. Additionally, kerosene showed an unexplained pressure drop deviation under specific conditions. Schlieben et al. [42] used liquid kerosene as the coolant to conduct experimental and numerical research on film cooling of GOX/kerosene rocket combustors, verifying that kerosene as a film coolant could adapt to various thermodynamic states such as liquid, transcritical, and supercritical.

Mist cooling is a promising technology where film cooling is enhanced with mist injection (such as water) [43, 44]. Unlike conventional film cooling, which employs continuous gas or liquid films, mist cooling introduces finely atomized liquid droplets into the flow field to enhance heat removal in high-temperature environments. These droplets function as localized cooling agents, evaporating in stages to absorb latent heat and create a cascading temperature reduction effect. The intermittent contact between droplets and the surface enables rapid energy extraction, while the transient evaporation process generates dynamic vapor layers that disrupt hot gas adhesion. This method extends effective cooling to downstream regions where single-phase coolant performance typically declines, making it particularly suitable for applications involving unsteady or complex thermal gradients. For example, Li et al. [45] conducted an investigation into the impingement of a mist/steam slot jet on a concave surface, with their simulation findings revealing that injecting water at 2% of the coolant flow rate could improve adiabatic cooling effectiveness by approximately 30–50%. Wang et al. [43] investigated mist cooling under gas turbine high pressure conditions (15 atm, 1561 K) and reported that mist cooling, implemented by injecting water droplets into the cooling air, enhanced film cooling performance. It was shown that introducing mist at a 10% mass fraction improved adiabatic cooling effectiveness by approximately 5–10%, and the greatest enhancement occurred when small droplets (about 5 μm) were used in combination with higher mist mass fractions.

Using liquid kerosene as the coolant for mist cooling takes advantage of both its latent heat of evaporation and its heat capacity, while also eliminating the need for additional cooling air. This is of considerable engineering significance for practical RDC systems that employ liquid hydrocarbon fuels as propellants. However, the use of kerosene-based mist cooling in RDCs has not yet been reported. The present study investigates the mist cooling characteristics of kerosene droplets injected through film holes on the outer wall surfaces of the RDC and compares their cooling performance with that of air film cooling and mist/air cooling. Here, air film cooling serves primarily as a baseline case, whereas mist or mist/air cooling is considered for scenarios where the availability of cooling air is limited, such as air-breathing RDC systems [46, 47]. Building on these results, an improved cooling strategy is proposed to enhance liquid film stability and heat removal capability, thereby providing guidance for the thermal protection design of liquid fuel RDC systems.

## 2. Numerical Methods

### 2.1 Physical Model and Boundary Conditions

The RDC model has an outer diameter of 36 mm, an inner diameter of 30 mm, and an axial height of 60 mm. Considering the symmetry of the RDC, four rows of film cooling holes are arranged circumferentially for simplicity and computational efficiency, with 10 holes per row and a diameter of $d = 0.6$ mm. The distance between the hole closest to the inlet and the inlet is $2.5d$, while the spacing between adjacent film holes is $5d$. The film holes occupy the upstream half of the axial length, as shown in Fig. 1. The cooling configurations will be discussed later. The RDC is ignited by a hot spot (2 MPa and 3000 K).

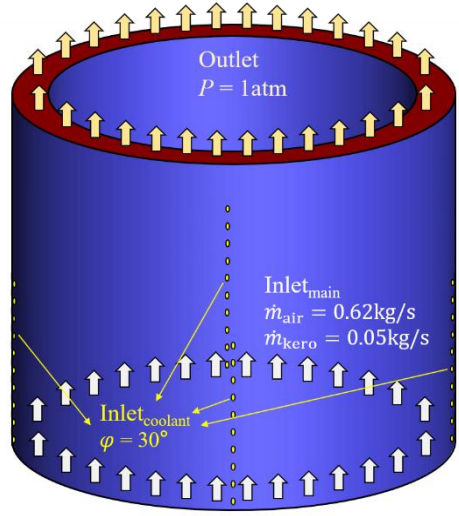

Figure 1. Film-cooled RDC model and boundary conditions.

The simulations are carried out using a density-based solver in ANSYS Fluent to solve the unsteady Reynolds-averaged Navier–Stokes (URANS) equations. Turbulence is modeled using the realizable k–ε approach with a finite-rate chemistry model. In accordance to Ref. [48], the standard wall function is used for near-wall treatment with no dedicated boundary-layer refinement applied; instead, the first-layer grid adjacent to the walls matches the uniform grid size of 0.3 mm used throughout the domain. This results in an effective y$^+$ range of approximately 5-20. Spatial discretization of both convection and diffusion terms is achieved using a second-order upwind scheme, and time integration is performed using a second-order implicit method. The kerosene droplets are tracked using the Discrete Phase Model (DPM):

$$\frac{d\boldsymbol{u}_d}{dt} = \frac{\boldsymbol{F}_d}{m_d} \quad (1)$$

$$\boldsymbol{F}_d = \frac{18\mu}{\rho_d D_d^2} \frac{C_d Re_d}{24} m_d |\boldsymbol{u} - \boldsymbol{u}_d| \quad (2)$$

where $\boldsymbol{u}_d, m_d, \rho_d$ and $D_d$ denote the droplet velocity vector, mass, density and diameter, respectively. $\boldsymbol{F}_d$ is the drag force on the spherical kerosene droplet. The droplet Reynolds number in Eq. (2) is defined as

$$Re_d = \frac{\rho D_d |\boldsymbol{u} - \boldsymbol{u}_d|}{\mu} \quad (3)$$

and the drag coefficient $C_d$ is calculated as

$$C_d = \begin{cases} 0.424, & Re_d > 1000 \\ \frac{24}{Re_d}\left(1 + \frac{1}{6}Re_d^{\frac{2}{3}}\right), & Re_d \leq 1000 \end{cases} \quad (4)$$

In the inert heating regime ($T_{\text{vap}} > T_d$), the droplet temperature $T_d$ is governed by a convective heat

balance:

$$m_d c_d \frac{dT_d}{dt} = hA_d(T_\infty - T_d) \quad (5)$$

When the droplet temperature reaches the specified vaporization temperature $T_{\text{vap}}$, the model transitions to the diffusion-controlled vaporization regime ($T_{\text{vap}} < T_d < T_{bp}$):

$$m_d c_d \frac{dT_d}{dt} = hA_d(T_\infty - T_d) - \frac{dm_p}{dt}h_{fg} \quad (6)$$

where $h_{fg}$ is the latent heat of vaporization and $T_{bp}$ is the boiling temperature. $c_p$ and $A_d$ represent the specific heat and surface area of the droplet. The convective heat transfer coefficient $h$ is calculated using the Ranz-Marshall relation [49, 50]. Radiative heat transfer is not considered, as particle-radiation interaction is not enabled in the present simulations. This assumption is considered reasonable for two-phase detonation flows, in which heat transfer between droplets and the surrounding gas is dominated by convection owing to large temperature gradients and high relative velocities.

A two-step mechanism for kerosene-air combustion proposed by Franzelli et al. [51] is used:

$$C_{10}H_{20} + 10O_2 \Rightarrow 10CO + 10H_2O \quad (7)$$

$$CO + 0.5O_2 \Leftrightarrow CO_2 \quad (8)$$

The chemical model parameters are given by Ref. [51]. The simplified chemical mechanism offers good computational efficiency, and its accuracy has been validated in previous studies [52-56].

The inlet boundary of the mainstream is set as a mass flow inlet, through which air and kerosene droplets are introduced respectively, and the direction is perpendicular to the inlet plane. The mass flow rate of air is 0.62 kg/s, the temperature is 1200 K, and the turbulence intensity is 5%. The kerosene droplets are injected with a mass flow rate of 0.05 kg/s, a temperature of 300 K, a droplet diameter of 5 μm, and an injection angle of 30° to the wall and inclined to the downstream. These inlet parameters are based on both experimental and numerical studies to ensure stable operation of a preheated air kerosene-fueled RDC under the present combustor geometry [47, 48, 52, 56, 57]. The outlet boundary is set as a pressure outlet of 1 atm to match the external ambient pressure. The RDC walls are set as adiabatic no-slip walls. In the mist-cooling configuration, kerosene droplets are injected into the RDC through the film-cooling holes, whereas in the air film-cooling configuration, ambient-temperature air is supplied as the coolant. A mixed mist/air film-cooling scheme is also considered, following Ref. [43], in which both kerosene droplets and cooling air are injected simultaneously. The parameters of the three cooling schemes will be discussed in detail in the subsequent sections.

## 2.2 Benchmark experiment validation

To verify the rationality of the film cooling related settings and the reliability of the simulation results presented in this study, a dedicated validation simulation is conducted using a flat plate film cooling model (Fig. 2). The configuration is adopted from a corresponding experimental study of Huo et al. [58]. The model settings are as follows: the mainstream inlet size is 30×80 mm, the length of the mainstream region is 320 mm, and the secondary flow is injected from film holes with a diameter $D$ of 4 mm and an inclination angle of 30°, and the distance between the adjacent holes is 16 mm. The computational grid is constructed with structured hexahedral meshes to ensure accuracy and quality. Local mesh refinement is applied to key regions, including the film cooling holes, the near-wall region of the flat plate, and the interaction zone between mainstream and secondary flow. The grid growth rate is controlled at 1.05 to prevent excessive distortion. The boundary layer mesh along the flat plate wall consists of 15 layers with a total thickness of 3 mm, and the first-layer grid height is optimized to maintain a $y^+$ range of 30-60, consistent with the standard wall-function approach used for near-wall turbulence modeling. The boundary conditions in the simulation are set according to the experimental conditions. The control temperature of the mainstream is 460 K, the flow rate is maintained at 0.08 kg/s, the temperature of air in the secondary flow is set to 345 K, and the temperature of water droplets is set to 300 K. The mass flow rate of the secondary flow is determined by the blowing ratio as follows:

$$\mathrm{M} = \frac{\rho_2 U_2}{\rho_1 U_1} = \frac{mf_2 A_1}{mf_1 A_2} \tag{9}$$

where $mf_1$ and $mf_2$ represent the mass flow rates of the mainstream and coolant respectively, $\rho_1$ and $U_1$ represent the mainstream density and flow velocity respectively, $\rho_2$ and $U_2$ represent the coolant density and flow velocity respectively, $A_1$ is the cross-sectional area of the mainstream inlet, and $A_2$ is the total outflow cross-sectional area of the film holes. The mass flow rate ratio of water droplets to coolant air is approximately 6%.

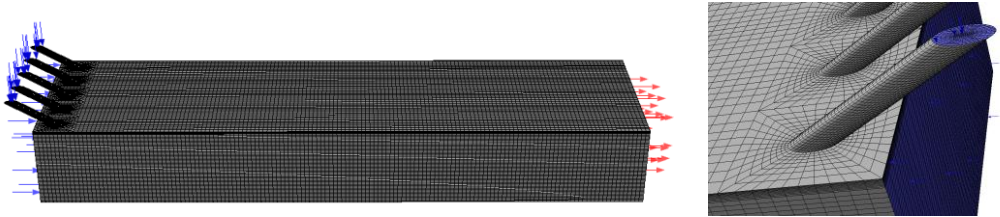

Figure 2. Computational setup of flat plate film cooling.

For each blowing ratio, simulations are performed using both air film cooling and mist/air cooling, and the results are compared with the experimental data [58]. The airflow direction is defined as the positive $x$-axis, and the film cooling effectiveness along the centerline of the single row of holes in the $x$ direction is evaluated accordingly. The comparisons show that the simulated cooling effectiveness follows the overall trend observed in the experiments, with minor local discrepancies. In addition to centerline cooling effectiveness, the normalized streamwise velocity profile in the wall-normal direction, $U/U_\infty$, is compared with experimental data, as shown in Fig. 3. For both $M = 0.5$ and $M = 1.0$, the simulations capture the key features: the near-wall low-velocity region, the velocity rise in the mixing layer, and the transition toward the mainstream flow. The overall agreement with experimental results is satisfactory, indicating that the present numerical approach can capture the main characteristics of the cooling jet and its interaction with the mainstream flow. However, it is noted that while the flat plate experiment indicates a relative improvement of approximately 3%-7% in cooling effectiveness for mist/air cooling over pure air cooling, the simulation predicts a larger enhancement in the range of 10%-20%, which may be attributed to limited fidelity and idealized simulation conditions.

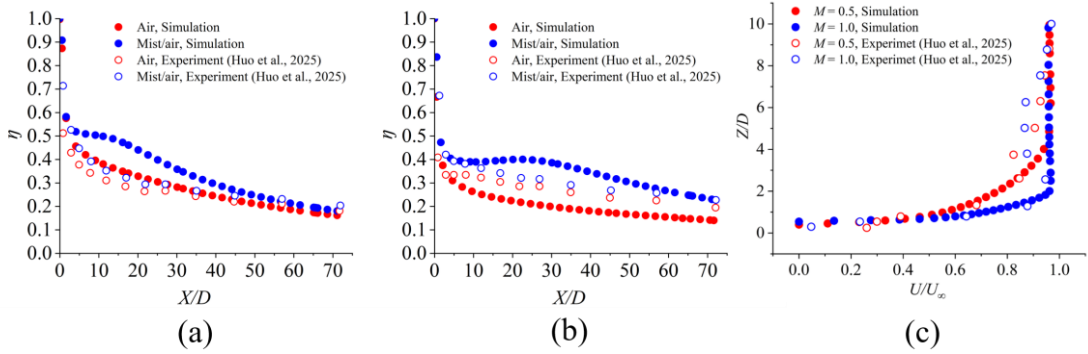

Figure 3. Comparison of flat plate cooling validation results between numerical simulation and experimental data [58]: (a) centerline cooling effectiveness at $M = 0.5$; (b) centerline cooling effectiveness at $M = 1.0$; (c) normalized streamwise velocity profile comparison.

It should be noted, however, that some discrepancies remain in the local mixing-layer region. These differences may stem from limitations of the turbulence model in capturing complex near-wall mixing, as well as simplifications in the numerical model regarding boundary conditions and geometric details. Nevertheless, these discrepancies are primarily confined to local features, while the overall distribution patterns and variation trends under different blowing ratios remain consistent with experimental observations. In addition to the flat-plate film cooling validation, the numerical framework for the detonation flow field has also been validated in our previous work [38] through comparisons between RDE simulations and experimental data.

**2.3 Grid sensitivity analysis**

A non-cooling benchmark case is evaluated using three grids with cell sizes of 0.25 mm, 0.30 mm, and 0.35 mm to perform a grid sensitivity analysis. The pressure profiles of the kerosene–air rotating detonation waves (RDWs) obtained from the three grid cases are shown in Fig. 4, and they nearly overlap with only minor differences. The deviations in RDWs' propagation velocity among the grids are within 1%, as indicated in the figure. For the pressure peak comparison, mean values are used due to inherent fluctuations. The peak pressures predicted by the 0.25 mm, 0.30 mm, and 0.35 mm grids are 4.55 MPa, 4.31 MPa, and 3.98 MPa, respectively, with only slight differences among them. The predicted detonation parameters are in close agreement with the velocity and pressure peaks calculated by NASA's CEA program [59] under the same initial conditions. Based on this analysis, the 0.30 mm grid is selected for the subsequent simulations.

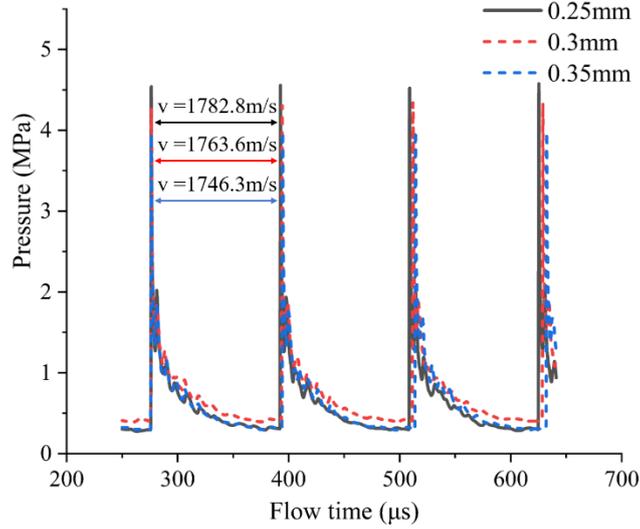

Figure 4. Pressure profiles of the kerosene-air RDWs under different grid resolutions.

## 3. Results and Discussion

### 3.1 Air film cooling analysis

The air film cooling case is presented first as a baseline for comparison, so that the enhancement achieved by kerosene mist injection can be evaluated more clearly. Four blowing ratios (0.5, 1.0, 2.0, and 3.0) are considered, as summarized in Table A-1 (Appendix A). The coolant inlet temperature is set to 300 K, and the corresponding mass flow rate is determined using the following definition of the blowing ratio:

$$M = \frac{\dot{m}_c}{A_c} \bigg/ \frac{\dot{m}_m}{A_m} \qquad (10)$$

Here, $\dot{m}_c$ is the coolant flow rate, and $A_c$ is the total area of the film holes; $\dot{m}_m$ is the mainstream total flow rate, $A_m$ is the mainstream cross-sectional area of the combustor. The mainstream total flow rate is the average value of the flow rate through the middle cross-section of the combustor within one detonation wave propagation cycle.

The snapshots of Fig. 5 illustrate the behavior of the cooling air issuing from the film holes when the RDW is located at angular positions of π/2, π, 3π/2 ahead of the holes, as well as immediately after the holes are swept by the detonation wave, corresponding to lines 1-4, respectively. At $M = 0.5$, the air coolant mass flow rate is relatively low, and distinct regions between adjacent film holes remain insufficiently covered by the coolant. The coolant film therefore does not fully isolate the mainstream flow from the wall. In contrast, at higher blowing ratios ($M = 2.0$ and 3.0), although the local cooling effectiveness at the hole exits is enhanced, pronounced discontinuities appear between adjacent coolant films, leading to deterioration of the overall cooling performance.

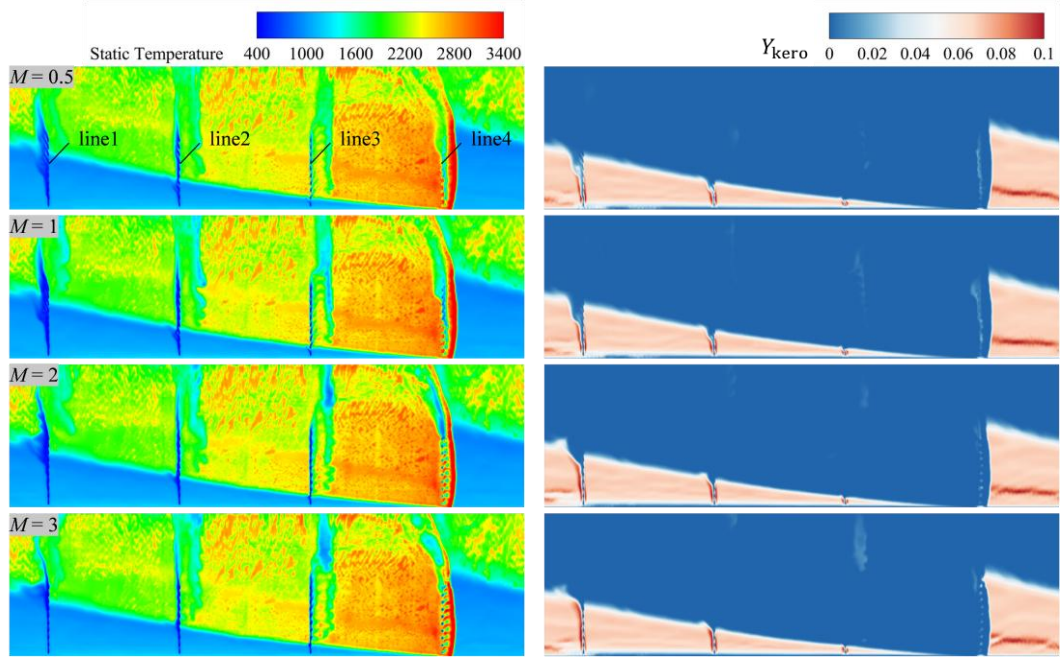

Figure 5. Rotating detonation flow fields under different blowing ratios of air cooling.

The instantaneous radial flow field during the passage of the RDW over the monitoring location (6000–6500 μs) is also analyzed in Fig. 6. The results indicate that variations in secondary flow injection pressure significantly influence the interaction between the mainstream flow and the cooling flow. For the case of $M = 0.5$, a pronounced reverse penetration of the mainstream into the upstream film holes is observed, and the downstream coverage of the cooling air is insufficient. At $M = 1.0$, the overall coverage of the cooling gas is noticeably improved; however, reverse penetration of the mainstream in the upstream region still occurs. At $M = 2.0$, the cooling air from all film holes forms a relatively continuous and stable film layer along the downstream wall. In contrast, at M = 3.0, the high injection pressure of the secondary flow causes the cooling gas to accumulate in the mainstream before the arrival of the high-pressure detonation wave. When the RDW sweeps over the film holes, the accumulated cooling gas is blown away from the wall, preventing the formation of a near-wall film cooling layer. Based on these observations, $M = 1.0$ and $M = 2.0$ are identified as the suitable secondary flow injection conditions among the four cases considered. An excessively low injection pressure leads to mainstream reverse penetration, which reduces the effective cooling film coverage and directly exposes the combustor wall to the high temperature mainstream. In addition, when high temperature gas flows back into the cooling holes, it directly heats the hole walls, potentially causing ablation, cracking, or structural failure if the material temperature limits are exceeded. The interaction between the reverse flow and the injected secondary flow also induces strong flow pulsations and pressure fluctuations inside the holes, further degrading the continuity and stability of the cooling film. Conversely, an excessively high injection pressure results in cooling gas accumulation, which strongly interferes with the mainstream flow and promotes the formation of large-scale recirculation zones or vortex structures, as shown in the figure.

The preceding discussion provides a qualitative analysis of the detonation flow field under air film cooling based on contour visualizations. To further compare the air film cooling performance at different blowing ratios, a quantitative analysis of the flow field is conducted in terms of the low-temperature region coverage, the temperature distribution along the hole centerline, the cooling effectiveness along the hole centerline, and the peak temperatures at different axial locations.

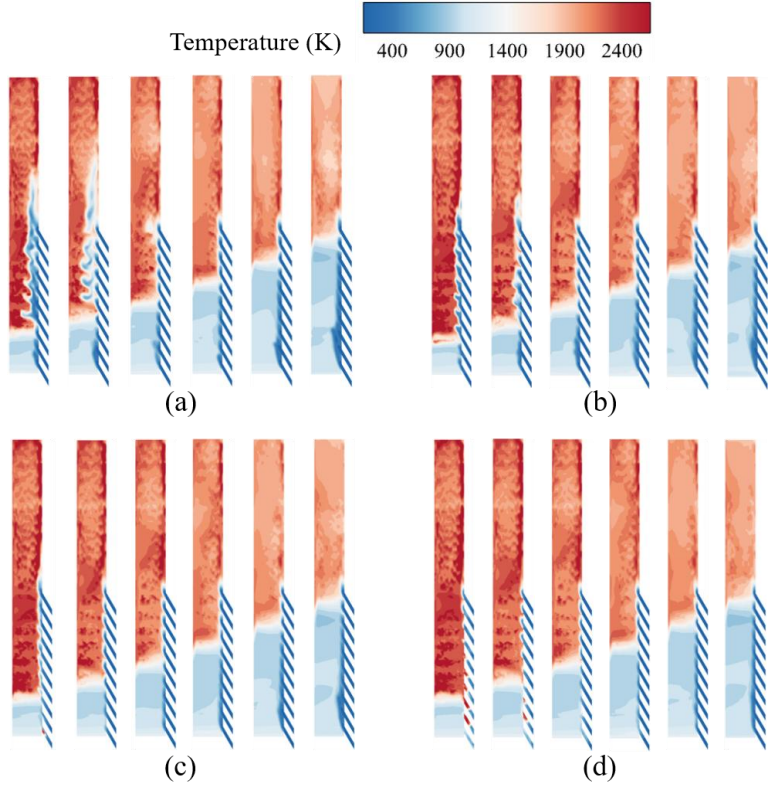

Figure 6. Instantaneous temperature contour fields near the film-cooling holes at different injection pressures during 6000–6500 μs: (a) $M = 0.5$, (b) $M = 1$, (c) $M = 2$, and (d) $M = 3$.

First, the low-temperature area coverage rate of the flow field near the outer wall is evaluated (Fig. 7). This metric is defined as the ratio of the area within 1 mm of the wall with a temperature less than or equal to a specified threshold temperature $T$ to the total computational area. The low temperature area coverage rates for air film cooling at different blowing ratios are presented in the figure below. The results show that the coverage rate in all temperature ranges is lowest at $M = 0.5$. As the blowing ratio increases, the low temperature area coverage rate initially increases; however, when the blowing ratio reaches $M = 3.0$, the coverage rate decreases. This trend indicates that, for air film cooling, an excessively high blowing ratio adversely affects the overall cooling performance. Temperature data are also extracted along the centerlines of the four rows of film holes, as shown in Fig. 8. The locations of these four lines are shown in Fig. 5. Lines 1 through 4 correspond to the moments when the detonation wave is located at π/2, π, and 3π/2 ahead of the holes, and when the holes have just been swept by the wave, respectively. Based on these sampling lines, the axial temperature variation curves for different blowing ratios at these moments are obtained and compared.

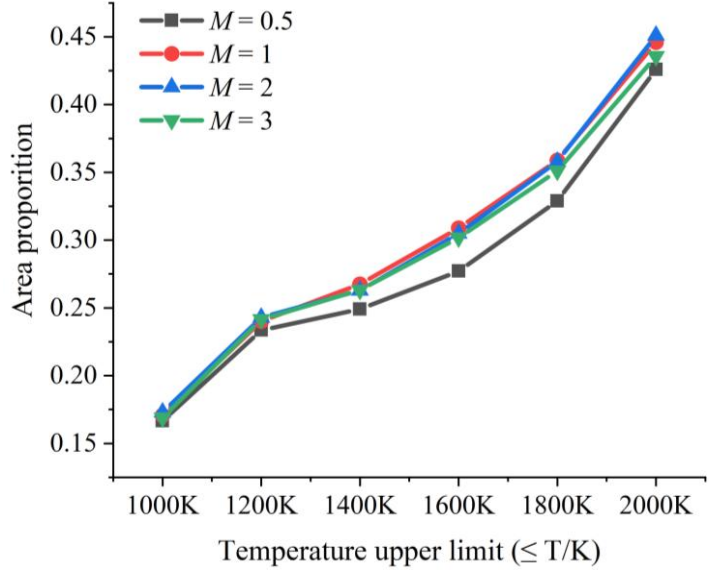

Figure 7. Coverage rate of the low-temperature region for air cooling.

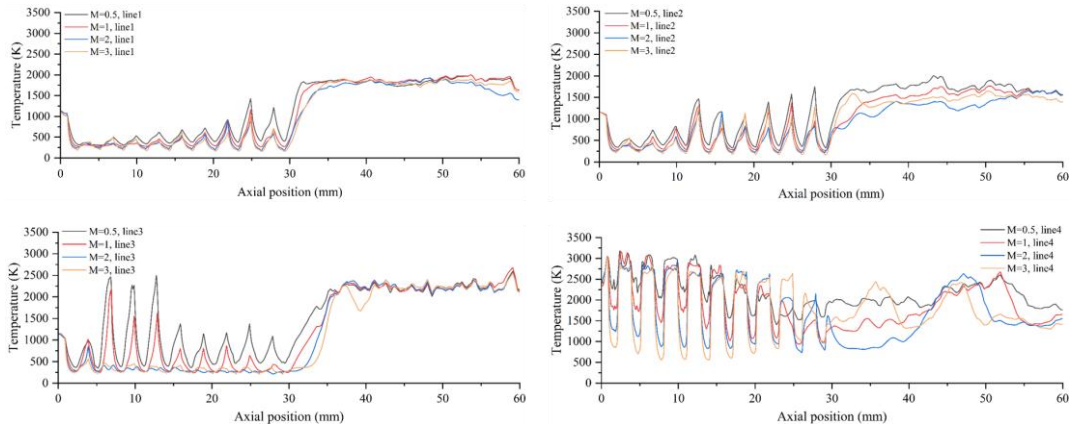

Figure 8. Temperature variation along the centerline of the air film holes.

The influence of the blowing ratio is primarily reflected in the temperature distributions near lines 3 and 4. As the blowing ratio increases from M = 0.5 to M = 2.0, the peak temperature decreases and the cooling performance improves. However, when the blowing ratio is further increased to M = 3.0, no significant improvement is observed; instead, a higher temperature peak appears at line 4. Owing to the inherently unsteady nature of the RDW, instantaneous results may be insufficient to fully characterize the cooling performance, and a comprehensive assessment over an entire detonation cycle is therefore required. In this study, monitoring points are placed along the centerline of the film holes at different axial locations. Figures 9 and 10 present the periodic temperature histories recorded at positions located one detonation wave height and one-half detonation wave height from the inlet on the inner wall of the combustor, respectively. Using the instantaneous simulation results without film cooling as the reference state t0, the corresponding film cooling efficiency is calculated. At the axial position corresponding to the height of the detonation wave, the overall film cooling efficiency increases with the blowing ratio from $M = 0.5$ to $M = 2.0$, but decreases when the blowing ratio reaches $M = 3.0$. At the axial position corresponding to half a detonation wave height, increasing the blowing ratio from $M = 0.5$ to $M = 1.0$ enhances the film cooling efficiency, whereas further increases from $M = 1.0$ to $M = 3.0$ result in a decline.

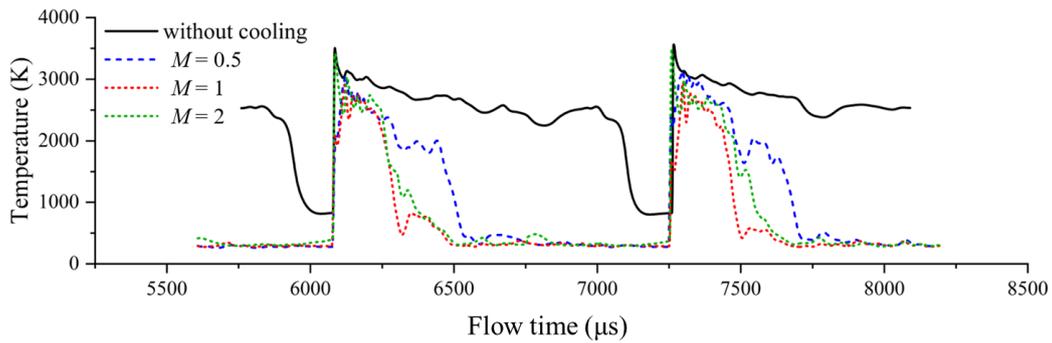

(a)

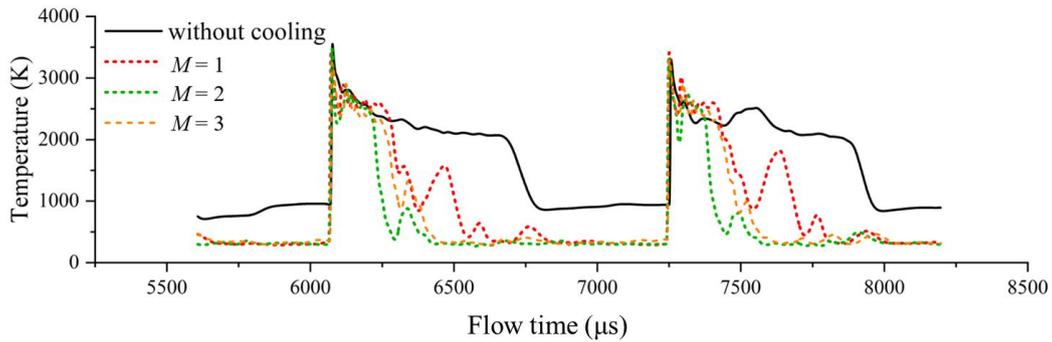

(b)

Figure 9. Temperature distributions at different vertical locations: (a) at the detonation wave height and (b) at half the detonation wave height.

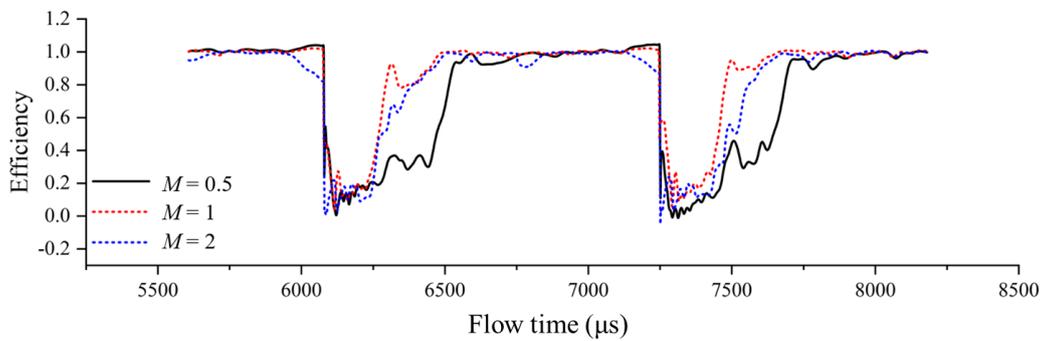

(a)

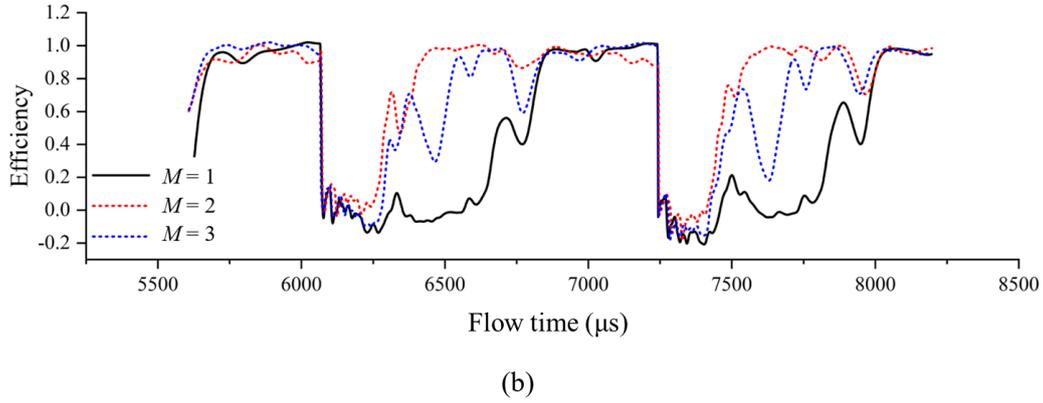

(b)

Figure 10. Cooling efficiency distributions at different vertical locations: (a) at the detonation wave height and (b) at half the detonation wave height.

The peak temperatures at different axial locations are recorded over two detonation cycles, and the corresponding results are presented in Fig. 11. As the blowing ratio increases from M=0.5 to M=1.0 and M=2.0, the temperature peaks at the axial monitoring points decrease. When the blowing ratio is further increased to $M$=3.0, the temperature peaks rise markedly at all monitoring locations. Overall, a moderate blowing ratio of $M$=1.0 or $M$= 2.0 represents the optimal compromise between cooling effectiveness, coverage extent, and film stability under the current configuration.

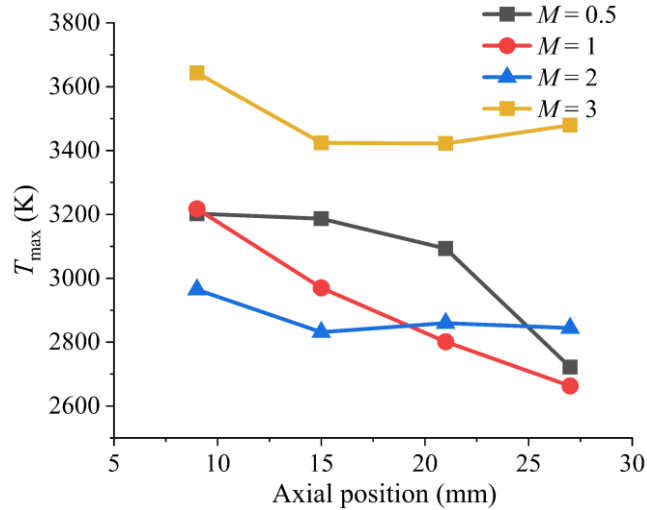

Figure 11. Temperature peaks at different axial positions under varying blowing ratios.

### 3.2 Kerosene droplet mist cooling analysis

In this configuration, kerosene droplets are injected through film holes along the wall and act as the coolant, forming a gas-liquid two-phase cooling film. Here we exploit the latent heat of phase change to achieve efficient heat absorption and significantly enhance wall thermal protection.

#### 3.2.1 Mist cooling with different blowing ratios

For comparison with air film cooling, four blowing ratios ($M$ = 0.5-3.0) are examined with a droplet diameter of 20 μm and temperature of 300 K. The resulting temperature and kerosene concentration fields are shown in Fig. 12. Temperature data are also extracted along the centerlines of the four rows of film holes, and axial temperature variation curves under different blowing ratios are obtained for comparison in Fig. 13.

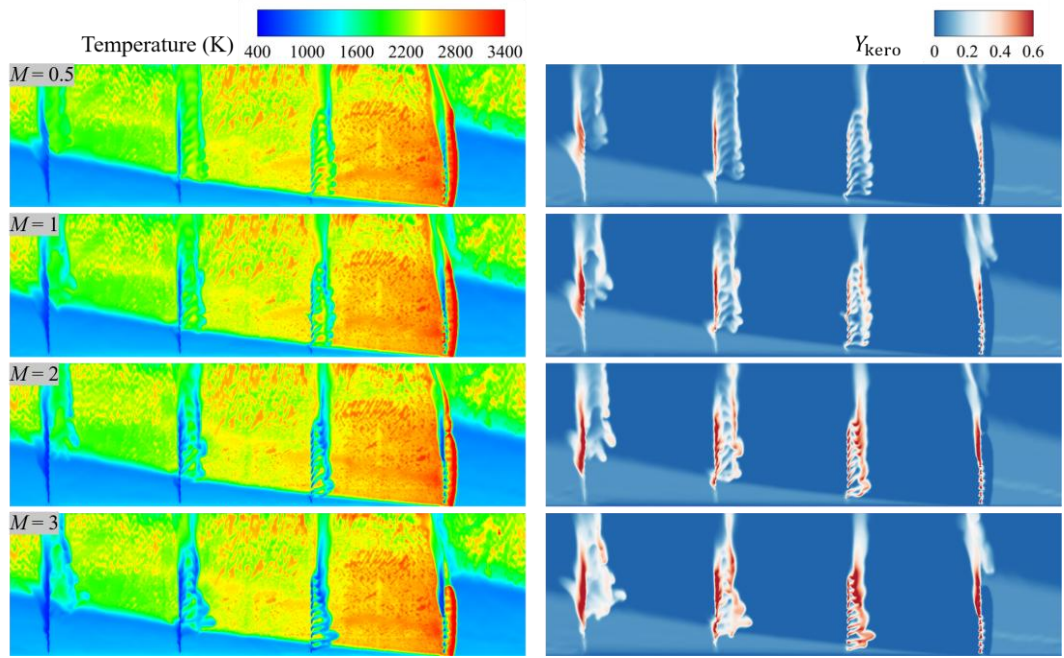

Figure 12. Temperature and kerosene concentration contours under different blowing ratios of kerosene droplet mist cooling.

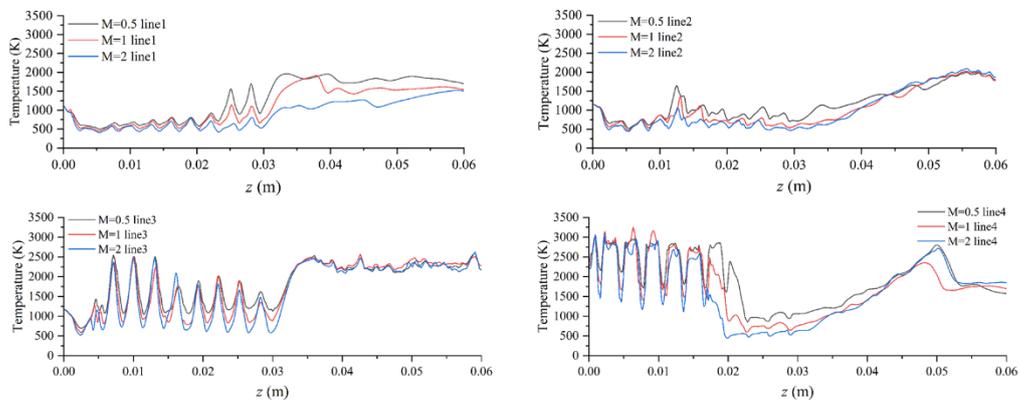

Figure 13. Temperature variation along the centerline of the mist cooling holes.

It can be observed that increasing the blowing ratio markedly enhances the near-wall film cooling effect, accompanied by a higher kerosene concentration in the cooling film. A key difference from air film cooling is that, for kerosene droplet mist cooling, the low-temperature area coverage (Fig. 14) increases monotonically with blowing ratio within the present range. Specifically, kerosene mist cooling promotes wall-adhering spreading and delays mixing with the mainstream flow. As a result, within a reasonable range, increasing the blowing ratio does not cause the liquid film to separate from the wall.

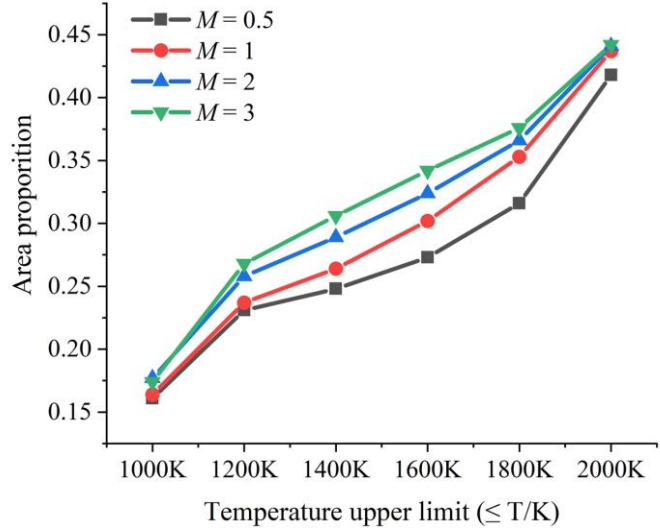

Figure 14. Coverage rate of the low-temperature region for kerosene mist cooling.

On the other hand, the contour plots show that kerosene droplet mist cooling has a pronounced influence on the detonation wave structure. Here, we evaluate its effect on the mainstream flow in terms of total pressure loss. The total pressure loss is evaluated based on the time-averaged total pressure ratio between the inlet and outlet planes of the combustor [60, 61]:

:

$$\varphi = \left(\frac{p_{t,\text{in}} - p_{t,\text{out}}}{p_{t,\text{in}}}\right) \quad (11)$$

where $p_{t,\text{in}}$ is the total pressure at the combustor inlet, and $p_{t,\text{out}}$ is the total pressure at the combustor outlet. The results indicate that as the blowing ratio increases from 0.5 to 3.0, the mainstream pressure loss increases to 13.8%, 13.4%, 14.2%, and 16.6%, respectively. Considering cooling effectiveness, film coverage, and pressure loss in combination, a moderate blowing ratio of $M$ = 1.0 or $M$ = 2.0 appears to provide a favorable compromise, achieving a balance between thermal protection performance and mainstream flow preservation. It should be noted that, for a total pressure gain achievement, the inlet and RDC configurations need to be carefully optimized [62], which is beyond the scope of the current study.

Additionally, to qualitatively assess the effects of air film cooling and kerosene mist cooling on the propagation of the detonation wave, the propagation velocity of the detonation wave is calculated for blowing ratios of 0.5, 1.0, 2.0, and 3.0. The results show that the average detonation wave velocity is 1763.6 m/s without cooling, 1770.7 m/s with air film cooling, and 1764.6 m/s with kerosene mist cooling. This indicates that similar to air film cooling [32, 35], the mist cooling has minimal impact on the propagation of the RDW, despite its influence on the mainstream and RDW structure.

**3.2.2 Mist cooling with different droplet sizes**

The second category examines operating conditions with different droplet diameters. All other parameters are kept constant, with the blowing ratio fixed at 1.0 and the droplet temperature set to 300 K, while only the kerosene droplet diameter is varied. Three droplet diameters are considered: 10 μm, 20 μm, and 50 μm. In selecting the kerosene droplet diameter, it is observed that when the diameter is smaller than 10 μm, combustion tends to occur at the outlets of the film holes. Accordingly, the droplet diameters considered in this study are limited to 10 μm, 20 μm, and 50 μm. The temperature and kerosene mass fraction contours of the near-wall flow field for different droplet diameters are shown in Fig. 15. The results indicate that the droplet diameter has a relatively minor overall effect on the cooling performance; however, under the condition with 50 μm droplets, the kerosene mass fraction near line 4 is noticeably lower.

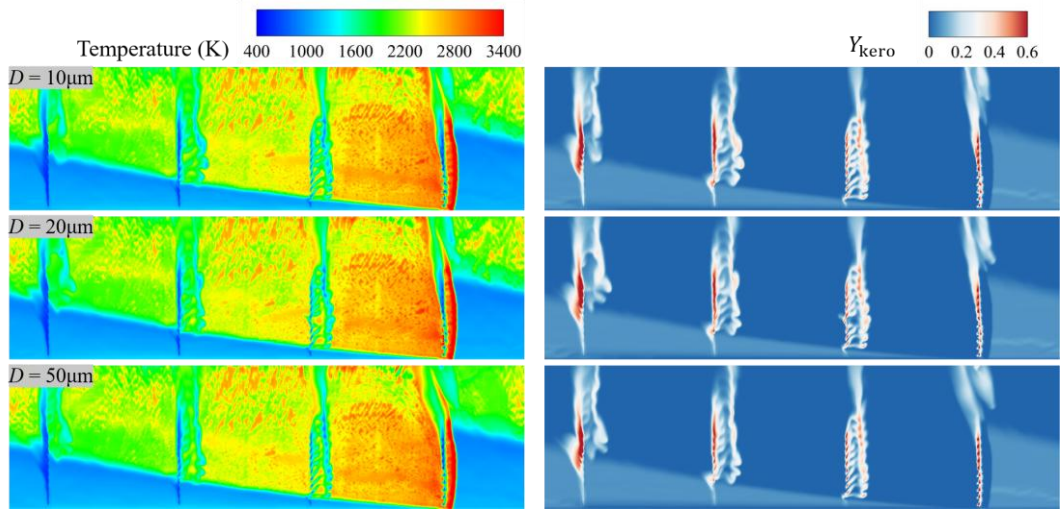

Figure 15. Temperature and kerosene concentration contours under different droplet diameters using kerosene droplet mist cooling.

From the axial temperature variation curves along the centerlines of the four rows of film holes in Fig. 16, it is evident that the kerosene droplet diameter has little influence on the cooling performance at lines 1, 2, and 3. At line 4, however, the downstream cooling effectiveness associated with 50 μm droplets is noticeably poorer. In conjunction with Fig. 15, this behavior can be attributed to the relatively slow evaporation rate of the 50 μm kerosene droplets when the detonation wave has just swept past the film holes. Owing to their smaller specific surface area, these larger droplets provide limited latent heat absorption in the immediate downstream region, resulting in reduced cooling effectiveness. As the flow evolves downstream, the droplets gradually evaporate and eventually achieve a cooling performance comparable to that of the 10 μm and 20 μm droplets. These results indicate that when kerosene droplets are used as the coolant, the droplet diameter should be neither too small nor too large. Based on this consideration, a droplet diameter of 20 μm is adopted for the remaining cases examined in this study.

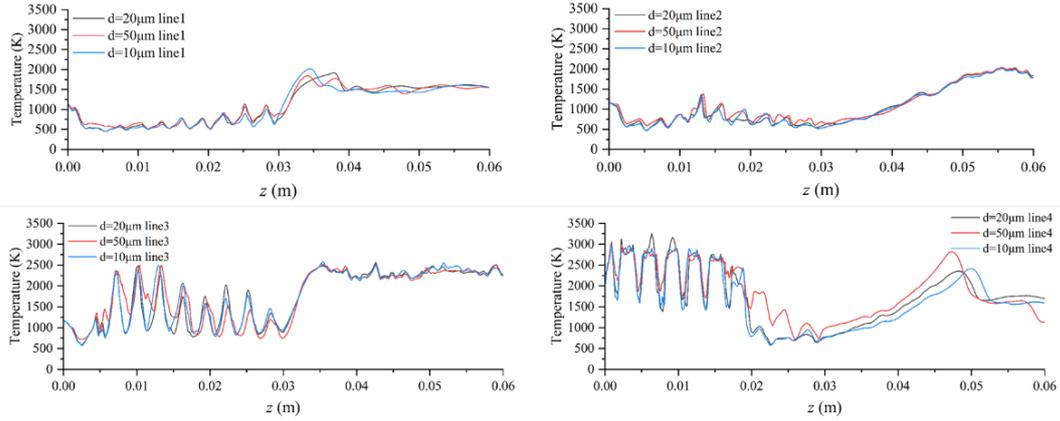

Figure 16. Temperature variation curves along the centerline of mist cooling film holes.

### 3.2.3 Mist/air cooling with different mass ratios

The preceding sections have identified the optimal parameters for single air film cooling and single kerosene droplet mist cooling. For air film cooling, a continuous and stable cooling film can be established at blowing ratios of $M$ = 1.0-2.0, achieving a balance between cooling effectiveness and flow field stability. For kerosene droplet mist cooling, a blowing ratio of $M$ = 1.0 combined with a droplet diameter of 20 μm provides enhanced heat flux control through latent heat absorption during phase change. To further explore the potential of synergistic cooling, a combined mist/air cooling strategy employing both air and kerosene droplets is investigated. The mass flow rate of coolants is 11.2 g/s, with a liquid kerosene mass fraction of 10%, and a droplet diameter of 20 μm.

Comparative cases without cooling and with air film cooling at $M$ = 1.0 are examined. The temperature evolution and cooling effectiveness at key wall monitoring locations, corresponding to one detonation wave height and one-half detonation wave height, are analyzed. As shown by the temperature histories at the monitoring points in Fig. 17, in the absence of film cooling, the peak wall temperature after passage of the detonation wave generally exceeds 2900 K, followed by a slow decay. In the later stage, the flow field temperature remains above 2000 K for an extended period. With air film cooling, the wall temperature is rapidly reduced to approximately 500 K. In comparison, the mist/air cooling further accelerates the cooling process. Compared to air cooling, the time required to reduce the temperature to around 300 K is shortened by approximately 10% in the detonation wave head region and by about 50% in the mid height region of the detonation wave.

On the basis that air film cooling has already reached its optimal blowing ratio ($M$ = 1.0), the introduction of a small amount of 20 μm kerosene droplets provides a significant supplementary cooling effect. Owing to inertia and surface tension, the droplets are able to maintain their integrity and are less sensitive to transient variations in the flow field. Even if some droplets deviate from their nominal trajectories, their dispersed distribution can still cover most of the wall surface, thereby reducing the risk of localized cooling failure. As a result, the mist/air cooling strategy achieves simultaneous improvements in cooling efficiency and film stability.

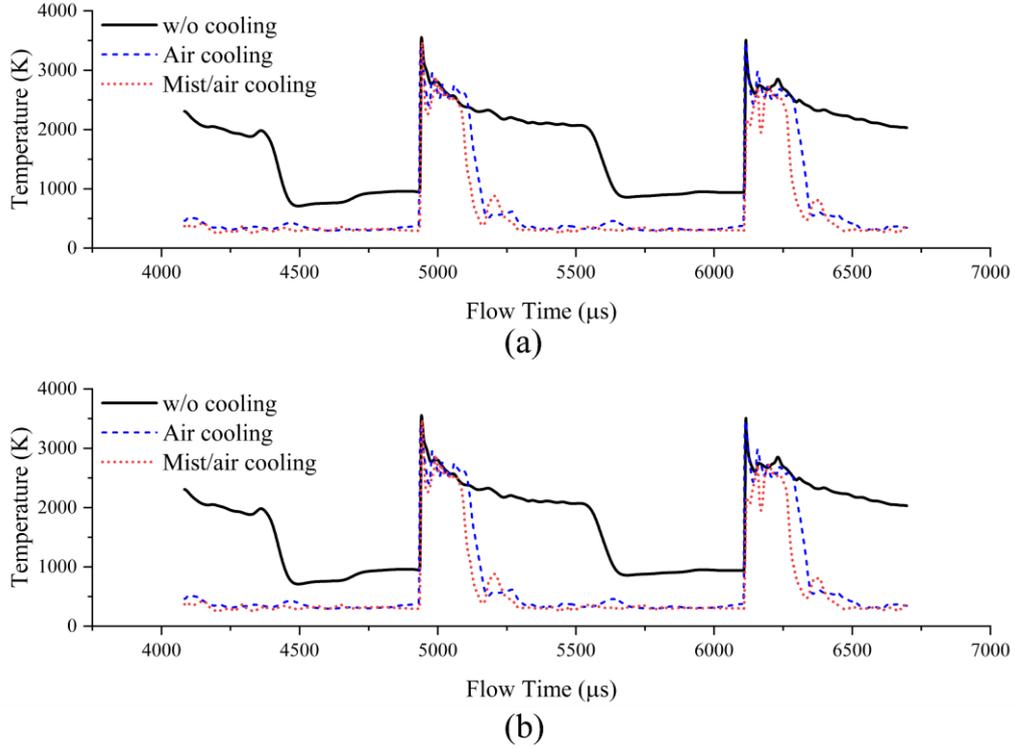

Figure 17. Temperature variation curves at different vertical locations: (a) at the detonation wave height and (b) at half-detonation wave height.

Because liquid kerosene also serves as a fuel, the combustion characteristics of the three cooling strategies are further examined. In this study, kerosene air combustion is described using a two-step chemical reaction model, as given in Eqs. (7) and (8), where CO is the product of the first step reaction and $CO_2$ is the product of the second step reaction. It can be observed from Figs. 18-20 that, compared with the case without cooling, the flow field with cooling exhibits a distinct banded region downstream of the secondary flow injection. This region becomes more pronounced closer to the inner wall. Within this banded area, the mass fractions of $H_2O$ and CO are relatively high, whereas the mass fraction of $CO_2$ is nearly zero. The spatial distribution of gaseous species in the contours reflects the combustion behavior of kerosene mist within the film-cooling flow. High concentrations of CO and $H_2O$ are confined to downstream banded regions originating from the film cooling holes and remain far from the primary thermal protection region of the RDC wall near the hole inlets. This indicates that kerosene mist combustion occurs only locally in non-critical cooling regions, avoiding excessive heat release to the protected wall. This behavior can be explained as follows. When the film cooling air is injected along the wall at high velocity, it induces strong turbulent vortices in the main combustion region, increasing the contact area between fuel and air and thereby promoting the first step reaction. Meanwhile, the secondary flow injection disrupts the originally more uniform airflow organization, effectively displacing CO, the product of the first step reaction, from oxygen-rich regions and transporting it downstream toward the inner wall. As a result,

CO accumulates in the banded region downstream of the secondary flow, forming a locally fuel rich zone that is unfavorable for the second step oxidation to $CO_2$. The differences among the three cooling strategies can be summarized as follows. For air film cooling, the injected air does not directly participate in combustion during the injection and diffusion stages of the secondary flow, but mainly redistributes the gas inside the combustor toward the downstream region and the inner wall. For mist/air film cooling, part of the kerosene droplets evaporates and participates in the first step reaction during the diffusion stage, producing CO and $H_2O$, which are subsequently transported downstream and toward the inner wall by the secondary flow. In contrast, kerosene droplet mist cooling produces the most pronounced banded structure, because a portion of the kerosene droplets evaporates and undergoes the first step reaction during both the injection and diffusion stages, leading to stronger local enrichment effects near the wall.

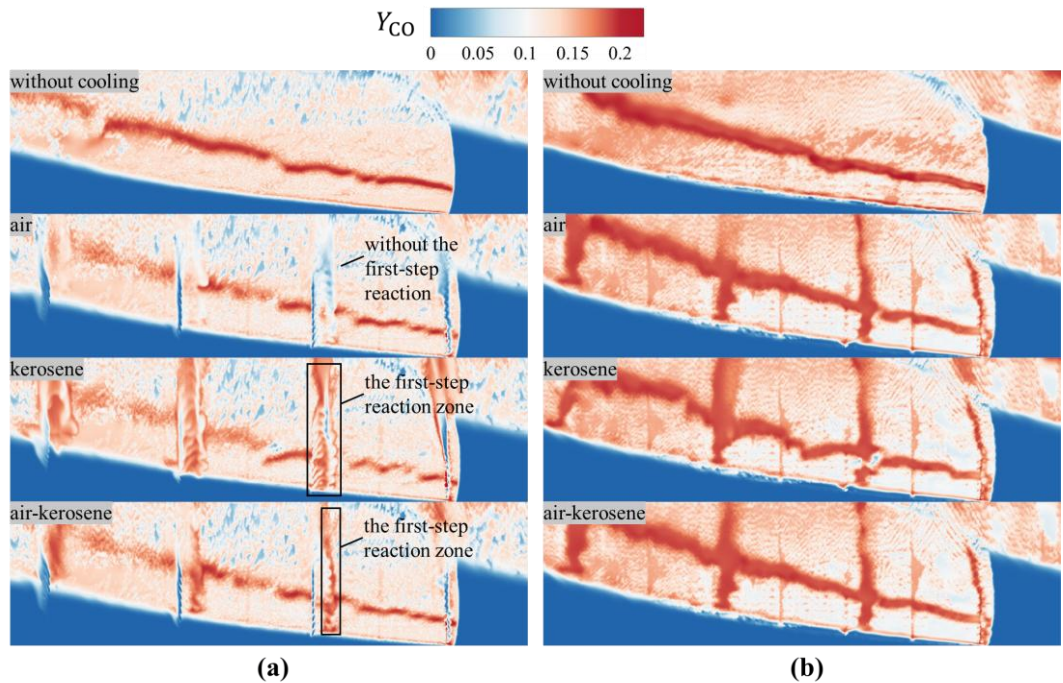

Figure 18. CO mass fraction distributions near the (a) outer wall and (b) inner wall.

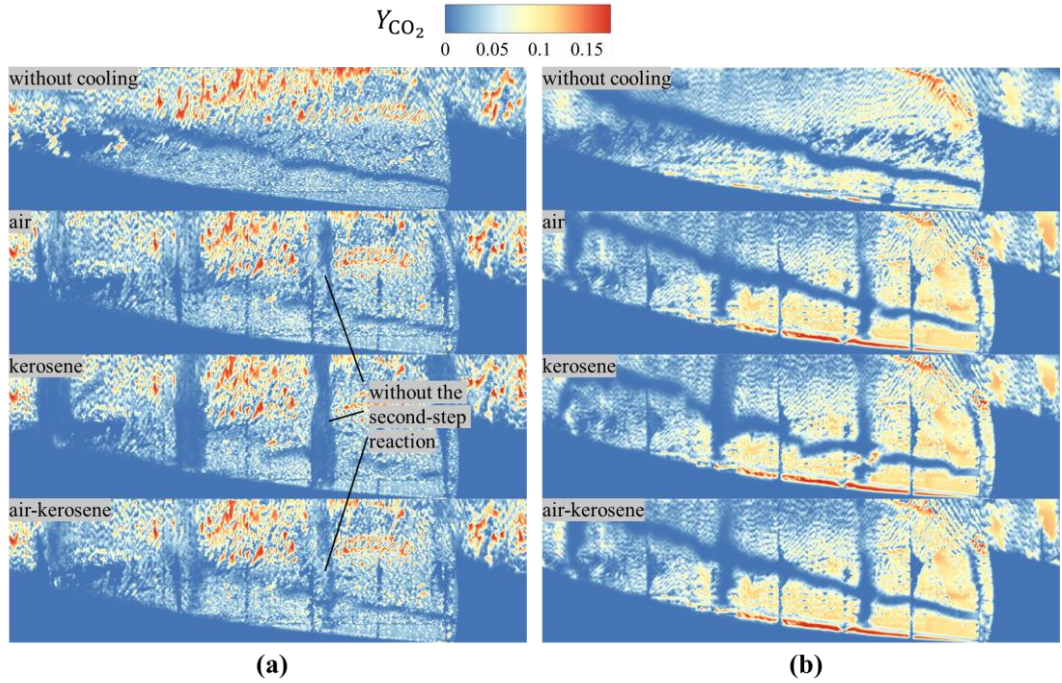

Figure 19. CO$_2$ mass fraction distributions near the (a) outer wall and (b) inner wall.

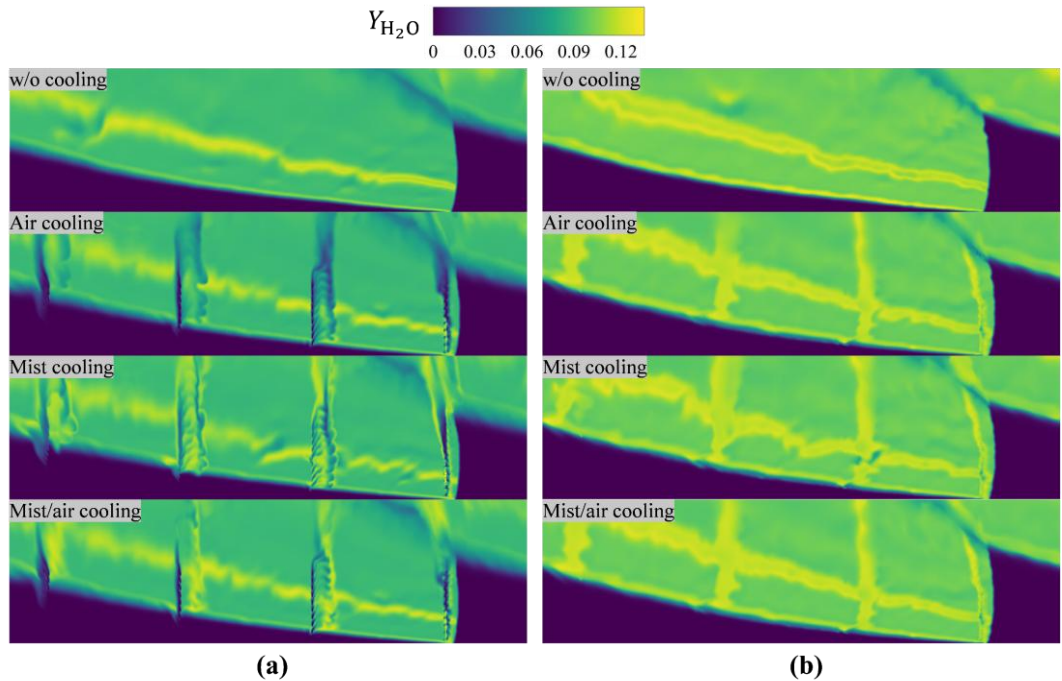

Figure 20. H$_2$O mass fraction distributions near the (a) outer wall and (b) inner wall.

The injection of kerosene droplets with the secondary flow increases the contact area between kerosene and the oxidizer, thereby enhancing the extent of the first-step reaction. As a result, kerosene droplet mist cooling exhibits a higher degree of first step reaction than mist/air cooling. However, the rapid progression of the first step reaction leads to a substantial depletion of oxygen, causing the second-step reaction to become limited by insufficient oxidizer availability. In addition, kerosene evaporation absorbs a large amount of heat, reducing the local temperature at the droplet-air interface and further suppressing the rate of the second step reaction. Figures 18–20 show that the $CO_2$ mass fraction in the near-wall flow field remains close to zero under all three cooling schemes. This indicates that the second-step, strongly exothermic stage of kerosene combustion is largely suppressed near the wall, and heat release is primarily limited to the milder initial reaction. As a result, the thermal load from kerosene combustion in the near-wall region is much smaller than the cooling effect provided by droplet phase change and the protective film layer. Moreover, the higher kerosene consumption rate is accompanied by greater reaction heat release. The gas species mass fraction contours also indicate that cooling strategies involving kerosene droplets undergo combustion during the secondary flow injection process. This implies that film cooling in these cases does not act on the original detonation combustion flow field alone, but rather on a flow field with an elevated thermal load due to additional reactions. Despite this, kerosene droplet mist cooling and mist/air cooling still demonstrate effective cooling performance. These results confirm the feasibility of incorporating kerosene droplets into film cooling strategies for combustor flow fields operating under liquid kerosene air detonation conditions.

## 4. Conclusions

This study investigates kerosene-based mist cooling as a wall thermal protection strategy for the RDC. An experimental flat-plate film-cooling configuration is first employed to validate the numerical framework and mist cooling models. The predicted cooling effectiveness shows good agreement with the trends observed in the experiments across a range of blowing ratios; however, it is also noted that the simulation may overpredict the cooling benefit due to limited predictive fidelity, and further experimental or high-fidelity validation is required to more conclusively assess the effectiveness of the mist cooling approach. On this basis, comparative numerical simulations of air film cooling, kerosene droplet mist cooling, and mist/air cooling are conducted in an RDC environment. The main findings are summarized as follows:

(1) The cooling effectiveness of air film cooling exhibits strong sensitivity to the blowing ratio. At low blowing ratios ($M = 0.5$), insufficient coolant momentum leads to incomplete film coverage, reverse penetration of hot gases into the film holes, and elevated near-wall temperatures, thereby reducing the effective thermal protection range by approximately 30% compared with higher blowing-ratio cases. Increasing the blowing ratio to $M = 1.0–2.0$ significantly improves film continuity and suppresses peak near-wall temperatures throughout an entire detonation cycle. However, at $M = 3.0$, excessive coolant injection destabilizes the cooling film due to strong interaction with the rotating detonation wave, resulting in film separation, increased temperature non-uniformity, and elevated peak near-wall temperatures at multiple axial locations.

(2) Kerosene droplet mist film cooling tends to provide stronger and more robust thermal protection than air film cooling. Under the same secondary mass flow rate, kerosene mist cooling achieves larger low-temperature surface coverage and generally lower near-wall peak temperatures, with the low-temperature coverage increasing monotonically with blowing ratio over the investigated range. This enhanced robustness is attributed to droplet inertia, wall-adhering behavior, and latent heat absorption during evaporation, which delay mixing with the mainstream flow and mitigate film separation. However, increasing the blowing ratio also intensifies the interaction between the mist film and the detonation wave, leading to higher mainstream total pressure losses, which increase from approximately 12% at $M$ = 0.5 to more than 15% at $M$ = 3.0. When cooling effectiveness and pressure loss are considered simultaneously, $M$ = 1.0-2.0 appears to be the preferred operating range for mist cooling. Within the investigated parameter space, droplet diameter exerts a limited influence on far-downstream cooling performance but significantly affects the near-injection region. Large droplets (50 μm) evaporate too slowly to provide effective cooling immediately downstream of the film holes, resulting in noticeably higher near-hole temperatures, whereas very small droplets tend to react prematurely near the injection location.

(3) Introducing a small mass fraction of kerosene droplets into an air film cooling configuration operating at its optimal blowing ratio ($M$ = 1.0) further enhances cooling performance. With a kerosene mass fraction of approximately 10% in the secondary flow, the mist/air scheme reduces the time required for the wall temperature to decrease to near-ambient levels by about 10% near the detonation wave head and by up to 50% at mid-wave height, compared with pure air film cooling. Relative to air-only cooling, the combined approach accelerates post-wave wall temperature recovery, improves cooling uniformity, and reduces sensitivity to local flow disturbances, while avoiding the larger pressure penalties associated with high-blowing-ratio mist cooling and limiting additional fuel consumption. In comparison, air film cooling does not directly participate in combustion, whereas kerosene-containing cooling schemes promote localized early-stage fuel oxidation near the wall, producing CO- and $H_2O$-rich regions and delaying $CO_2$ formation. Despite the associated chemical heat release, both mist-only and combined mist and air cooling schemes achieve a net reduction in wall thermal load.

(4) Finally, to further validate the results and confirm the effects of mist cooling, experiments conducted under identical or closely matched conditions to the numerical model are required. Such experiments would provide direct evidence of the cooling performance and wall protection in realistic detonation flow conditions. In addition, high-fidelity simulations are needed to more accurately capture the unsteady interactions between the rotating detonation flow. Conducting these efforts in future work will yield a more comprehensive understanding of kerosene mist film cooling in two-phase RDCs.

**Appendix A: Simulation cases summary**

Table A presents a summary for all operating conditions considered in this study, including four cases of air film cooling, six cases of kerosene mist cooling, and one case of combined kerosene mist and air cooling. All data are averaged over more than five complete RDW cycles to eliminate transient fluctuations in the rotating detonation flow field, ensuring that the results are representative. The average cooling effectiveness refers to the overall mean cooling efficiency along the centerlines of the four rows of film-cooling holes.

Table A-1 Summary of simulation cases and average cooling effectiveness

| Cooling Scheme | $M/D$ (μm) | Peak Wall Temperature (K) | Total Pressure Loss (%) | Average Cooling Effectiveness (%) | Low-Temperature (1600K) Coverage Rate (%) |
|---|---|---|---|---|---|
| Air | 0.5/20 | 3222.5 | 14.9 | 13.8 | 27.7 |
| Air | 1.0/20 | 3217.5 | 13.9 | 23.2 | 30.8 |
| Air | 2.0/20 | 2964.4 | 13.0 | 27.9 | 30.5 |
| Air | 3.0/20 | 3643.3 | 11.9 | 25.9 | 30.2 |
| Kerosene | 0.5/20 | 3084.5 | 13.8 | 14.9 | 27.3 |
| Kerosene | 1.0/10 | 3122.0 | 13.2 | 26.0 | 27.0 |
| Kerosene | 1.0/20 | 3062.0 | 13.4 | 24.4 | 30.2 |
| Kerosene | 1.0/50 | 3107.6 | 14.0 | 18.0 | 26.8 |
| Kerosene | 2.0/20 | 2964.5 | 14.2 | 28.0 | 32.4 |
| Kerosene | 3.0/20 | 2873.6 | 16.6 | 35.0 | 34.2 |
| Air/Kerosene | 1.0/20 | 3040.6 | 14.6 | 25.2 | 30.2 |